\documentclass[final,5p,times,twocolumn]{elsarticle}
\usepackage{graphicx,color,epsfig,rotate,amssymb}

\journal{Solid State Communications}

\begin{document}

\begin{frontmatter}

\title{Study of ground state phases for spin-1/2 Falicov-Kimball model on a triangular lattice}

\author[iitr]{Sant Kumar}
\author[iisc]{Umesh K. Yadav}
\author[iitr]{T. Maitra}
\author[iitr]{Ishwar Singh}
\address[iitr]{Department of Physics, Indian Institute of Technology Roorkee, Roorkee- 247667, Uttarakhand, India}
\address[iisc]{Department of Physics, Centre for Condensed Matter Theory, Indian Institute of Science, Bangalore-560012, India}

\begin{abstract}
The spin-dependent Falicov-Kimball model (FKM) is studied on a triangular lattice using numerical diagonalization technique and Monte-Carlo simulation algorithm. Magnetic properties have been explored for different values of parameters: on-site Coulomb correlation $U$, exchange interaction $J$ and filling of electrons. We have found that the ground state configurations exhibit long range Ne\`el order, ferromagnetism or a mixture of both as $J$ is varied. The magnetic moments of itinerant ($d$) and localized ($f$) electrons are also studied. For the one-fourth filling case we found no magnetic moment from $d$- and $f$-electrons for $U$ less than a critical value.
\end{abstract}


\end{frontmatter}

\section{Introduction}

The correlated electron systems like cobaltates~\cite{qian06,tera97,tekada03}, $GdI_{2}$~\cite{tara08} and its doped variant $GdI_{2}H_{x}$ ~\cite{tulika06,fel,ryaz}, $NaTiO_{2}$~\cite{clarke98,pen97,khom05}, $MgV_{2}O_{4}$~\cite{rmn13} etc. have attracted great interest recently as they exhibit a number of remarkable cooperative phenomena such as valence and metal-insulator transition, charge, orbital and spin/magnetic order, excitonic instability and possible non-fermi liquid states~\cite{tara08}. These are layered triangular lattice systems and are characterized by the presence of localized (denoted by $f$-) and itinerant (denoted by $d$-) electrons. The geometrical frustration from the underlying triangular lattice coupled with strong quantum fluctuations give rise to a huge degeneracy at low temperatures resulting in competing ground states close by in energy. Therefore, for these systems one would expect a fairly complex ground state magnetic phase diagram and the presence of soft local modes strongly coupled with the itinerant electrons. It has recently been proposed that these systems may very well be described by different variants of the two-dimensional Falicov-Kimball model (FKM)~\cite{tara08,tulika06} on the triangular lattice. 

Originally the FKM was proposed to describe the metal-insulator transition in mixed valence compounds~\cite{fkm69,fkm70}. 
Later the FKM was used to study the tendency of formation of charge density wave (CDW) 
order as  well~{\cite{brandt,schmidt,freer,hassan}. Recently we have studied the ground state and finite temperature properties of the FKM and its different extensions on the triangular lattice~\cite{umesh1,umesh2}. We have reported several interesting results like various charge order, metal-insulator transitions and resolved the issue of spontaneous symmetry breaking (SSB)~\cite{umesh2} in the ground state and explored the metal-insulator transition at finite-temperature~\cite{umesh3,umesh4,umesh5} in the different regime of parameters. In all these studies the spin-degree of freedom was ignored and the interactions between electrons were spin-independent. Recent experimental results show that a charge order generally occurs with an attendant spin/magnetic order in many correlated systems ~\cite{chen,tranq1,tranq2}. In order to describe both the charge and magnetic orders in a unified way we use a generalized FKM Hamiltonian~\cite{lemanski05} that includes spin-dependent local interactions:

\begin{eqnarray}
H=-\,\sum\limits_{\langle ij\rangle\sigma}(t_{ij}+\mu\delta_{ij})d^{\dagger}_{i\sigma}d_{j\sigma}
+\,(U-J)\sum\limits_{i\sigma}f^{\dagger}_{i\sigma}f_{i\sigma}d^{\dagger}_{i\sigma}d_{i\sigma}
\nonumber \\
+\,U\sum_{i\sigma}f^{\dagger}_{i,-\sigma}f_{i,-\sigma}d^{\dagger}_{i\sigma}d_{i\sigma}
+\,U_{f}\sum\limits_{i\sigma}f^{\dagger}_{i\sigma}f_{i\sigma}f^{\dagger}_{i,-\sigma}f_{i,-\sigma}
\nonumber \\
+E_{f}\sum\limits_{i\sigma}f^{\dagger}_{i\sigma}f_{i\sigma}
\end{eqnarray}
\noindent here $\langle ij\rangle$ denotes the nearest neighbor ($NN$) lattice sites. The $d^{\dagger}_{i\sigma},  d_{i\sigma}\,(f^{\dagger}_{i\sigma},f_{i\sigma})$ are, respectively, the creation and annihilation operators for $d$- ($f$-) electrons with spin $\sigma=\{\uparrow,\downarrow\}$ at the site $i$. First term is the band energy of the $d$-electrons and $\mu$ is the chemical potential. The hopping parameter $t_{\langle ij\rangle} = t$ for $NN$ hopping and zero otherwise. The interaction between $d$-electrons is neglected in FKM as usual. The second term is the on-site interaction between $d$ and $f$-electrons of same spin with coupling strength ($U - J$) (where $U$ is the usual spin-independent Coulomb term and $J$ is the exchange interaction; the term follows from Hund's coupling). The third term is the on-site interaction $U$ between $d$- and $f$-electrons of opposite spins.  { Here $J$ basically represents the spin dependent local interactions between localized ($f$-) and itinerant ($d$-) electrons that stabilizes parallel over anti-parallel alignment between $f$- and $d$-electrons. Inclusion of the exchange or Hund's coupling term enables us to study the magnetic structure of the $f$-electrons and band magnetism of the $d$-electrons. Fourth term is on-site Coulomb repulsion $U_f$ between opposite $f$-spins while the last term is the spin-independent, dispersionless energy level$E_f$ of the $f$-electrons.

There are some theoretical results available for the spin-dependent FKM on a bipartite lattice ~\cite{lemanski05,farkov02}. A few ground state charge and magnetic configurations exist for certain fixed values of $U$ and $J$. There is hardly any study available for the spin-dependent FKM on non-bipartite lattices. Therefore, in the present work we take up model systems that represent layered materials with triangular lattice (hence geometrically frustrated). Within second order perturbation theory, the spinless FKM with extended interactions  can be shown to map to an effective Ising model with antiferrmagnetic interactions in the large U limit~\cite{umesh2}. The AFM coupling on triangular lattice is frustrated and leads to large degeneracies at low temperature. It turns out that this frustration is lifted~\cite{gruber1,gruber2} in the higher order perturbation in $\frac{1}{U}$~\cite{footnote}. Therefore it would be quite interesting to see the role of spin degree of freedom of electrons on the ground state properties on such lattices with different values of parameters $U$ and $J$. We study FKM at different range of interactions $U$ and $J$ for different electronic filling fractions on a triangular lattice. 

\section{Methodology}

All the interactions in the Hamiltonian $H$ (Eq.$1$) preserve local occupation and spin of the $f$-electrons, i.e. the $d$-electrons traveling through the lattice change neither occupation numbers nor spins of the $f$-electrons. The local $f$-elctron occupation number $\hat{n}_{fi\sigma}=f_{i\sigma}^{\dagger}f_{i\sigma}$ is conserved as $\big[\hat{n}_{fi\sigma},H\big]=0$ for all $i$ and $\sigma$. This implies that $\omega_{i\sigma}=f_{i\sigma}^{\dagger}f_{i\sigma}$ is a good quantum number taking values only $1$ or $0$ as the site $i$ is occupied or unoccupied by an $f$-electron of spin $\sigma$, respectively. Following this local conservation, $H$ can be rewritten as 
\begin{eqnarray}
H=\sum\limits_{\langle ij \rangle \sigma}\, h_{ij}(\{\omega_{\sigma}\})\,d_{i\sigma}^{\dagger}d_{j\sigma}
+\,U_{f}\sum\limits_{i\sigma}{\omega_{i\sigma}\omega_{i,-\sigma}}
\nonumber \\
+\,E_{f}\sum\limits_{i\sigma}\,{\omega_{i\sigma}}
\end{eqnarray}
\noindent where $h_{ij}(\{\omega_{\sigma}\})=\big[-t_{ij}+\{(U-J)\omega_{i\sigma}+U\omega_{i,-\sigma}-\mu\}\delta_{ij}\big]$ and $\{\omega_{\sigma}\}$ is a chosen configuration of $f$-electrons of spin $\sigma$.

The Hamiltonian $H$ in Eq.$2$ shows that the $f$-electrons act as an external charge and spin-dependent potential or annealed disordered background for the non-interacting $d$-electrons. This external potential of $f$-electrons can be ``annealed'' to find the minimum energy of the system. It is clear that there is inter-link between subsystems of $f$- and $d$-electrons. This inter-link is responsible for the long range ordered configurations and different charge and magnetic structures of $f$-electrons in the ground state.

We set the scale of energy with $t_{\langle ij \rangle} = 1$. The value of $\mu$ is chosen such that the filling is ${\frac{(N_{f}~ + ~N_{d})}{4N}}$ (e.g. $N_{f} + N_{d} = N$ is one-fourth case and $N_{f} + N_{d} = 2N$ is half-filled case etc.), where $N_{f} = (N_{f_{\uparrow}}+N_{f_{\downarrow}})$, $N_{d} = (N_{d_{\uparrow}} + N_{d_{\downarrow}})$ and $N$ are the total number of $f-$ electrons, $d-$ electrons and sites respectively. For a lattice of $N$ sites the $H(\{\omega_{\sigma}\})$ (given in Eq.2) is a $2N\times 2N$ matrix for a fixed configuration $\{\omega_{\sigma}\}$. For one particular value of $N_f(= N_{f_{\uparrow}} + N_{f_{\downarrow}})$, we choose values of $N_{f_{\uparrow}}$ and $N_{f_{\downarrow}}$ and their configuration $\{\omega_{\uparrow}\} = \{{\omega_{1\uparrow}, \omega_{2\uparrow},\ldots, \omega_{N\uparrow}}\}$ and $\{\omega_{\downarrow}\} = \{{\omega_{1\downarrow}, \omega_{2\downarrow},\ldots, \omega_{N\downarrow}}\}$. Choosing the parameter $U$ and $J$, the eigenvalues $\lambda_{i\sigma}$($i = 1\ldots N$) of $h(\{\omega_{\sigma}\})$ are calculated using the numerical diagonalization technique on the triangular lattice of finite size $N(=L^{2}, L = 12)$ with periodic boundary conditions (PBC).

The partition function of the system is written as,
\begin{eqnarray}
\it{Z}=\,\sum\limits_{\{\omega_{\sigma}\}}\,Tr\,\left(e^{-\beta H(\{\omega_{\sigma}\})}\right)
\end{eqnarray}
\noindent where the trace is taken over the $d-$electrons, $\beta=1/k_{B}T$. The trace is calculated from the eigenvalues $\lambda_{i\sigma}$ of the matrix $h(\{\omega_{\sigma}\})$ (first term in Eq.2). The partition function can, therefore, be recast in the form,
\begin{eqnarray}
\it{Z}=\,\sum\limits_{\{\omega_{\sigma}\}}\,
\prod\limits_{i}\,\left(e^{-\beta\big[U_{f}\omega_{i\sigma}\omega_{i,-\sigma} + E_{f}\omega_{i\sigma} \big]}\right)\,
\nonumber \\
\prod\limits_{j}\,\left(e^{-\beta\big[\lambda_{j\sigma}(\{\omega_{\sigma}\})-\mu\big]}+1 \right)
\end{eqnarray}
Now, the thermodynamic quantities can be calculated as averages over various configurations $\{\omega_{\sigma}\}$ with
statistical weight $P(\{\omega_{\sigma}\})$ is given by
\begin{eqnarray}
P(\{\omega_{\sigma}\})=\frac{e^{-\beta\,F(\{\omega_{\sigma}\})}}{\it{Z}}
\end{eqnarray}
\noindent where the corresponding free energy is given as,
\begin{eqnarray}
F(\{\omega_{\sigma}\})=\,-\frac{1}{\beta}\,\bigg[ln\,\left(\prod\limits_{i}\,
e^{-\beta\big[U_{f}\omega_{i\sigma}\omega_{i,-\sigma} + E_{f}\omega_{i\sigma} \big]}\right)
\nonumber \\
+\sum\limits_{j}\,ln\,\left(e^{-\beta\big[\lambda_{j\sigma}(\{\omega_{\sigma}\})-\mu\big]}+1 \right) \bigg]
\end{eqnarray}
\noindent The ground state total internal energy $E(\{\omega_{\sigma}\})$ is calculated as,
\begin{eqnarray}
E(\{\omega_{\sigma}\})=\,\lim_{T\rightarrow 0} F(\{\omega_{\sigma}\})
=\sum\limits_{i\sigma}^{N_{d}}\lambda_{i\sigma}(\{\omega_{\sigma}\})
\nonumber \\
+U_{f}\sum\limits_{i\sigma}\omega_{i\sigma}\omega_{i,-\sigma}
+E_{f}\sum\limits_{i\sigma}\omega_{i\sigma}
\end{eqnarray}

Our aim is to find the unique ground state configuration (state with minimum total internal energy $E(\{\omega_{\sigma}\}$)) of $f$-electrons out of exponentially large possible configurations for a chosen $N_{f}$. In order to achieve this goal, we have used classical Monte Carlo simulation algorithm by annealing the static classical variables $\{\omega_{\sigma}\}$ ramping the temperature down from a high value to a very low value. Details of the method can be found in our earlier
papers~\cite{umesh1,umesh2,umesh3,umesh4,umesh5,umesh6}.

\section{Results and discussion}

\begin{figure}[h]
\begin{center}
\includegraphics[trim=0.5mm 0.5mm 0.5mm 0.5mm,clip,width=8.9cm,height=7.0cm]{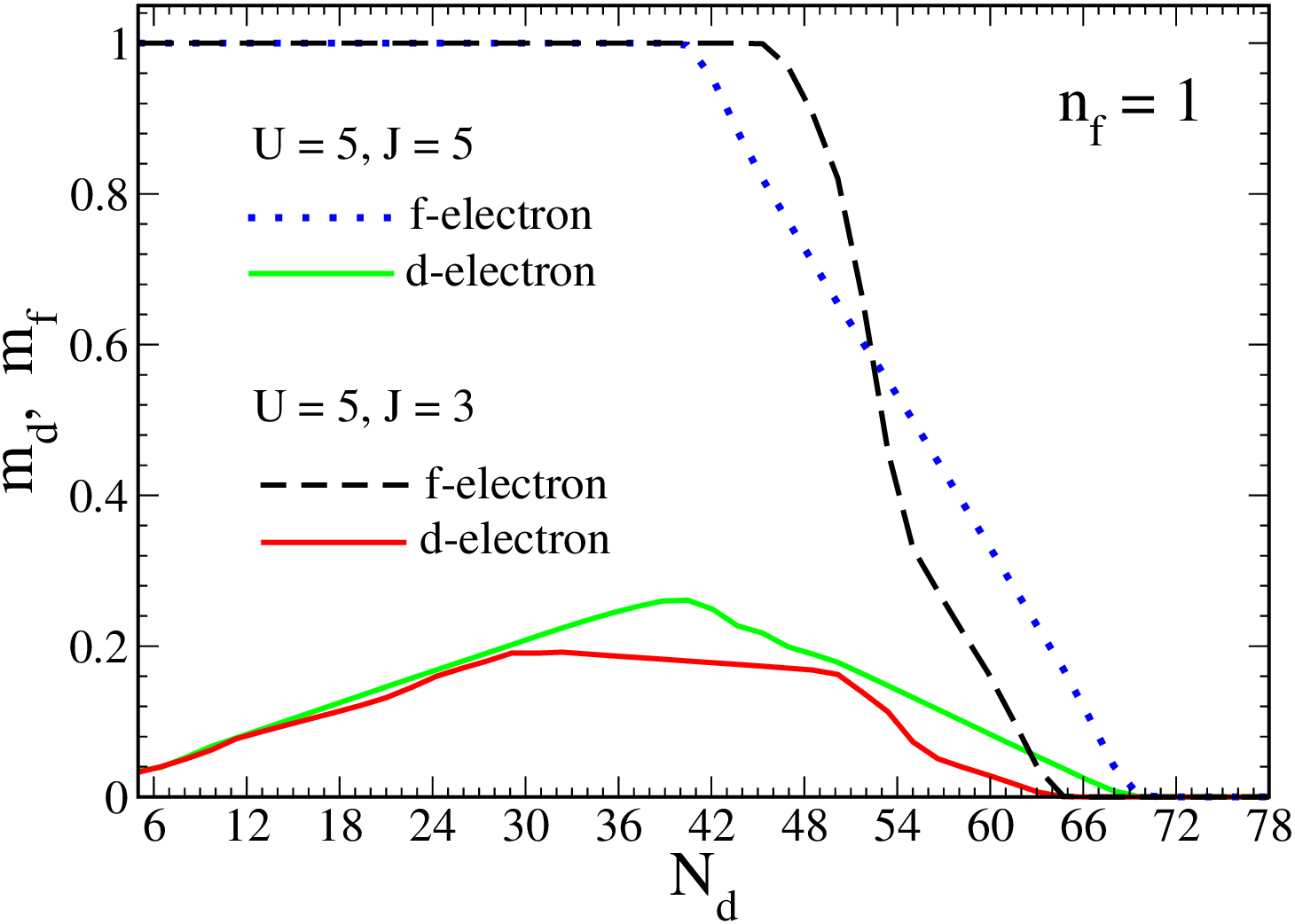}
\caption{(Color online) Variation of magnetic moment of $d$-electrons $m_{d}$ and $f$-electrons $m_{f}$ with number of $d-$electrons $N_d$ for $n_{f}=1$, $U = 5$, $U_{f} = 10$ and for $J = 5$ and $3$.}
\end{center}
\end{figure}

\begin{figure}[h]
\begin{center}
\includegraphics[trim=0.4mm 0.4mm 0.4mm 0.1mm,clip,width=8.9cm,height=4.0cm]{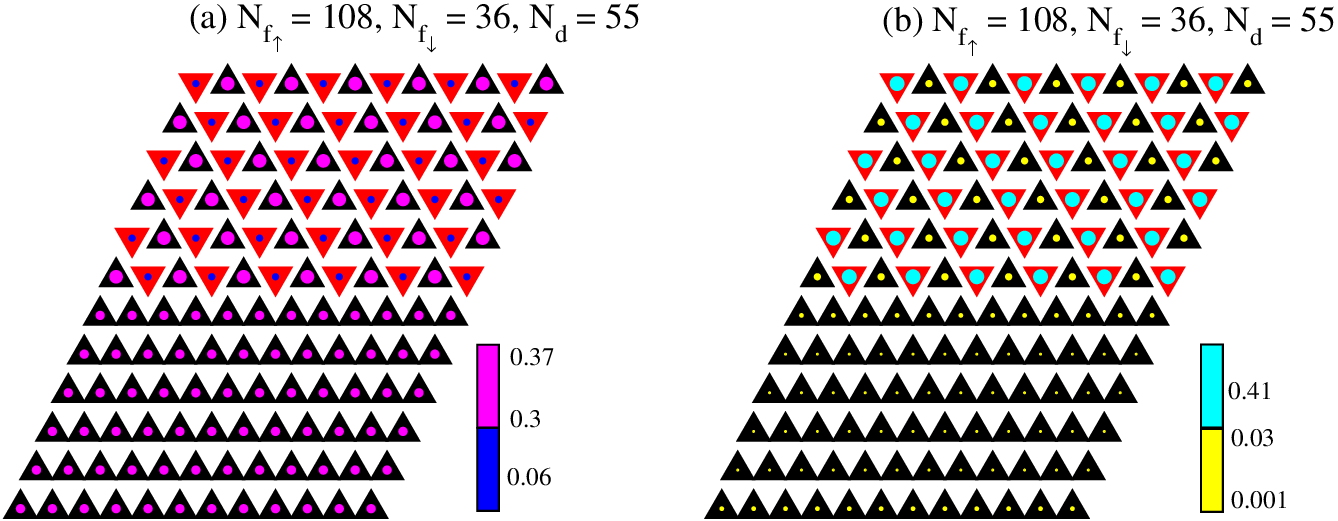}
\caption{(Color online) (a) Up-spin and (b) down-spin $d$-electron densities are shown on each site for $J = 5$, $U = 5$, $U_{f} = 10$, $n_{f} = 1$ and $N_{d} = 55$. The color coding and radii of the circles indicate the $d$-electron density profile. Triangle-up and triangle-down, filled by black and red colors correspond to the sites occupied by up-spin and down-spin $f$-electrons, respectively.}
\end{center}
\end{figure}

We have studied the variation of magnetic moment of $d$-electrons $\large(m_{d} = \frac{M_{d}}{N} = \frac{\large(N_{d_{\uparrow}}~ - ~N_{d_{\downarrow}}\large)}{N}\large)$ (as $d$-electrons are spread over all $N$ number of sites) and magnetic moment of
 $f-$electrons $\large(m_{f} = \frac{M_{f}}{N_f} =  \frac{\large(N_{f_{\uparrow}}~ - ~N_{f_{\downarrow}}\large)}{N_f}\large)$
 (as $f$-electrons are confined to only $N_f$ number of sites) with number of $d$-electrons $N_{d}$ at a fixed value of $U$, $U_{f}$ and $J$ for $n_{f} = 1~\large(n_{f} = \frac{N_{f}}{N} = \frac{\large(N_{f_{\uparrow}}~ + ~N_{f_{\downarrow}}\large)}{N}\large)$. We have also studied the density of $d$-electrons at each site for the above case. Fig.$1$ shows the variation of magnetic moment of $d$-electrons ($m_{d}$) and $f$-electrons ($m_{f}$) with number of $d$-electrons ($N_d$) for two different values of exchange correlation $J$ i.e. $J = 5$ and $J = 3$ at a fixed value of on site coulomb repulsion $U = 5$ and $U_{f} = 10$. From Fig.$1$ one can note that when $N_{d} = 144$ the ground state is Neel ordered anti-ferromagnetic (AFM) in nature. The reason for this can be understood in the following way. It is clear From Eq.$1$ that for $U = J$, there is no repulsion between $d$- and $f$-electrons of same spins on the same site. The repulsion between $d$- and $f$-electrons of the opposite spins on the same site is $U$. Therefore it is energetically favorable that a $d$-electron of same spin (as that of $f-$) occupies the site. For the FM arrangement of $f$-electrons, all $d$-electrons occupying the sites shall have FM arrangement themselves. Similarly for the AFM arrangement of $f$-electrons, $d$-electrons occupying the sites shall have AFM arrangement. For FM arrangements of $d$- and $f$-electrons, there is no hopping possible for $d$-electrons due to Pauli$^{\prime}$s exclusion principle, but for AFM arrangement of $f$- and $d$-electrons, there will be finite hopping of $d$-electrons between neighboring sites. This hopping reduces the kinetic energy of $d$-electrons and hence total band energy of $d$-electrons. Hence AFM arrangement of spins corresponds to minimum energy. Thus the ground state is AFM. In fact, it remains AFM up to $N_{d} = 70$ for $J = 5$ and up to $N_{d} = 65$ for $J = 3$. The magnetic moment for $f$-electrons starts increasing for $N_{d} < 70$ (for $J = 5$) and for $N_{d} < 65$ (for $J = 3$). Fully FM state is observed at $N_{d} \le 40$ for $J=5$ and at $N_{d} \le 45$ for $J = 3$, as now $d$-electrons find a plenty of sites (where no $d$-electrons are present) to hop to.
By changing the number of $d$-electrons, basically we are varying the doping and the extent of doping caused the phase transition from Neel order AFM to FM via mixture of both state. These type of phase transition also observed using band structure calculation for $GdI_{2}H_{x}$, by doping hydrogen as reported by authors~\cite{tulika06,fel,ryaz}.

For the above mentioned parameter value $U = 5$ and $J = 5$ and $U = 5$ and $J = 3$, we have seen $f$-electrons configuration and $d$-electrons density at each site for different $N_{d}$. Fig.$2$ shows the density of $d$-electrons at each site for $U = 5, J = 5$ for $N_d = 55$ (say). Table $1$ and $2$, respectively summaries the density of $d$-electrons for two cases $U = 5$ and $J = 5$ (Fig.$3$)  and $U = 5$ and $J = 3$ (not shown here) with FM and AFM arrangement of $f$-electrons. The density of $d$-electrons at each site strongly depends upon the value of exchange correlation $J$. Let us compare the density of $d$-electrons at sites with up-spin and down-spin $f$-electrons for a fixed value of $N_{d}$ (say $N_{d} = 55$) and for $J = 5$ and $J = 3$ (given in Table $1$ and $2$). We note from Table $1$ that the density of $d$-electrons at sites where $d$- and $f$-electrons have same spins is large for $J = 5$ and less for $J = 3$. This is expected (as seen from Hamiltonian (Eq.$1$)), because for $U = J = 5$, there is no on-site repulsion between $d$- and $f$-electrons of the same spins, but for $U = 5$ and $J = 3$ there is finite repulsion between $d$- and $f$-electrons of the same spins. Also from Table $1$ and $2$, we note that density of $d$-electrons with up-spin ($n_{d_{\uparrow}}$) at sites with FM arrangement of $f$-electrons is lesser than that at sites with AFM arrangement of $f$-electrons. This is so because in later case the $d$-electrons may hop to either empty sites or to sites having down-spin $f$-electrons. This hopping reduces the kinetic energy of $d$-electrons and hence total band energy of $d$-electrons. For the FM case, the $d$-electrons can hop only to empty sites and the reduction in its total band energy is less. Same is true for density of $d$-electrons with down-spin ($n_{d_{\downarrow}}$).

\begin{table*}
\caption{The density of $d$-electrons with FM arrangement of $f$-electrons}
\centering
\begin{tabular}{|*{2}{c|}*{2}{c|}c|}
\hline
\multicolumn{2}{|c|}{Density of $d$-electrons} & \multicolumn{2}{|c|}{Density of $d$-electrons}& $U = 5$ \\
\multicolumn{2}{|c|}{with up spin ($n_{d_{\uparrow}}$)} & \multicolumn{2}{|c|}{with down spin ($n_{d_{\downarrow}}$)}& $N_{d} = 55$  \\
\hline
Sites with $f_{\uparrow}$ & Sites with $f_{\downarrow}$ & Sites with $f_{\uparrow}$ & Sites with $f_{\downarrow}$ & \\
\hline
$0.3$ & 0 & $0.001 - 0.03$ & $0$ & $J = 5$ \\
\hline
$0.27$ & 0 & $0.058 - 0.072$ & $0$ & $J = 3$ \\
\hline
\end{tabular}
\end{table*}

\begin{table*}
\caption{The density of $d$-electrons with AFM arrangement of $f$-electrons}
\centering
\begin{tabular}{|*{2}{c|}*{2}{c|}c|}
\hline
\multicolumn{2}{|c|}{Density of $d$-electrons} & \multicolumn{2}{|c|}{Density of $d$-electrons} & $U = 5$ \\
\multicolumn{2}{|c|}{with up spin ($n_{d_{\uparrow}}$)} & \multicolumn{2}{|c|}{with down spin ($n_{d_{\downarrow}}$)}& $N_{d} = 55$  \\
\hline
Sites with $f_{\uparrow}$ & Sites with $f_{\downarrow}$ & Sites with $f_{\uparrow}$ & Sites with $f_{\downarrow}$ & \\
\hline
$0.37$ & $0.06$ & $0.06$ & $0.41$ &  $J = 5$ \\
\hline
$0.32$ & $0.1$ & $0.1$ & $0.3$ & $J = 3$ \\
\hline
\end{tabular}
\end{table*}

We have also studied the ground state magnetic phases for up-spin and down-spin $f$-electrons, magnetic moments of $d$- and $f$-electrons and the density of $d$-electrons on each site for the range of values of parameters $U$, $U_{f}$ and $J$ for two cases (i) $n_{f} + n_{d} = 1$ (one-fourth filled case) and (ii) $n_{f} + n_{d} = 2$ (half filled case). We have chosen large value of $U_{f}$ so that double occupancy of $f$-electrons is avoided.

\subsection{One-fourth filled case ($n_{f} + n_{d} = 1$):}

\begin{figure}[h]
\includegraphics[trim=0.4mm 0.4mm 0.4mm 0.1mm,clip,width=8.9cm,height=4.0cm]{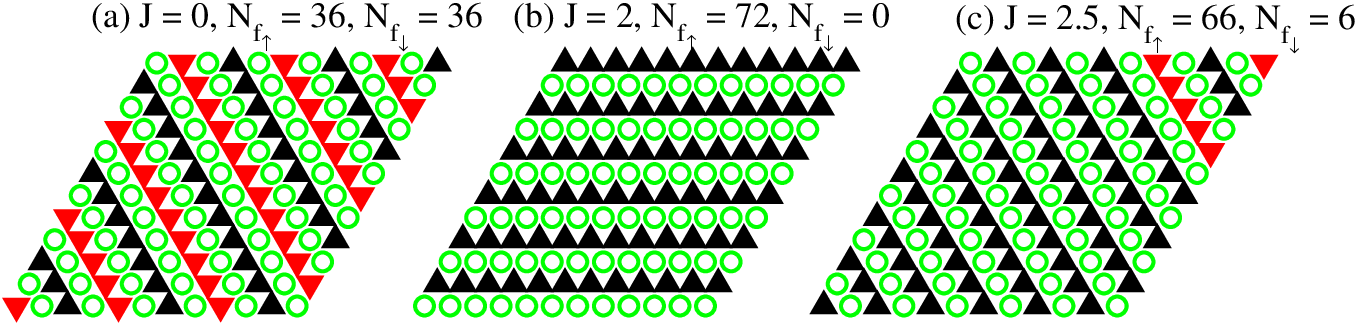}
\caption{(Color online) The ground-state magnetic configurations of $f$-electrons for $n_{f} = \frac{1}{2}$, $n_{d} = \frac{1}{2}$, $U = 5$, $U_{f} = 10$ and for various values of $J$. Triangle-up and triangle-down, filled by black and red colors correspond to the sites occupied by up-spin and down-spin $f$-electrons, respectively. Open green circles correspond to the unoccupied sites.}
\end{figure}

\begin{figure}[h]
\includegraphics[trim=0.5mm 0.0mm 0.5mm 0.1mm,clip,width=8.9cm,height=5.5cm]{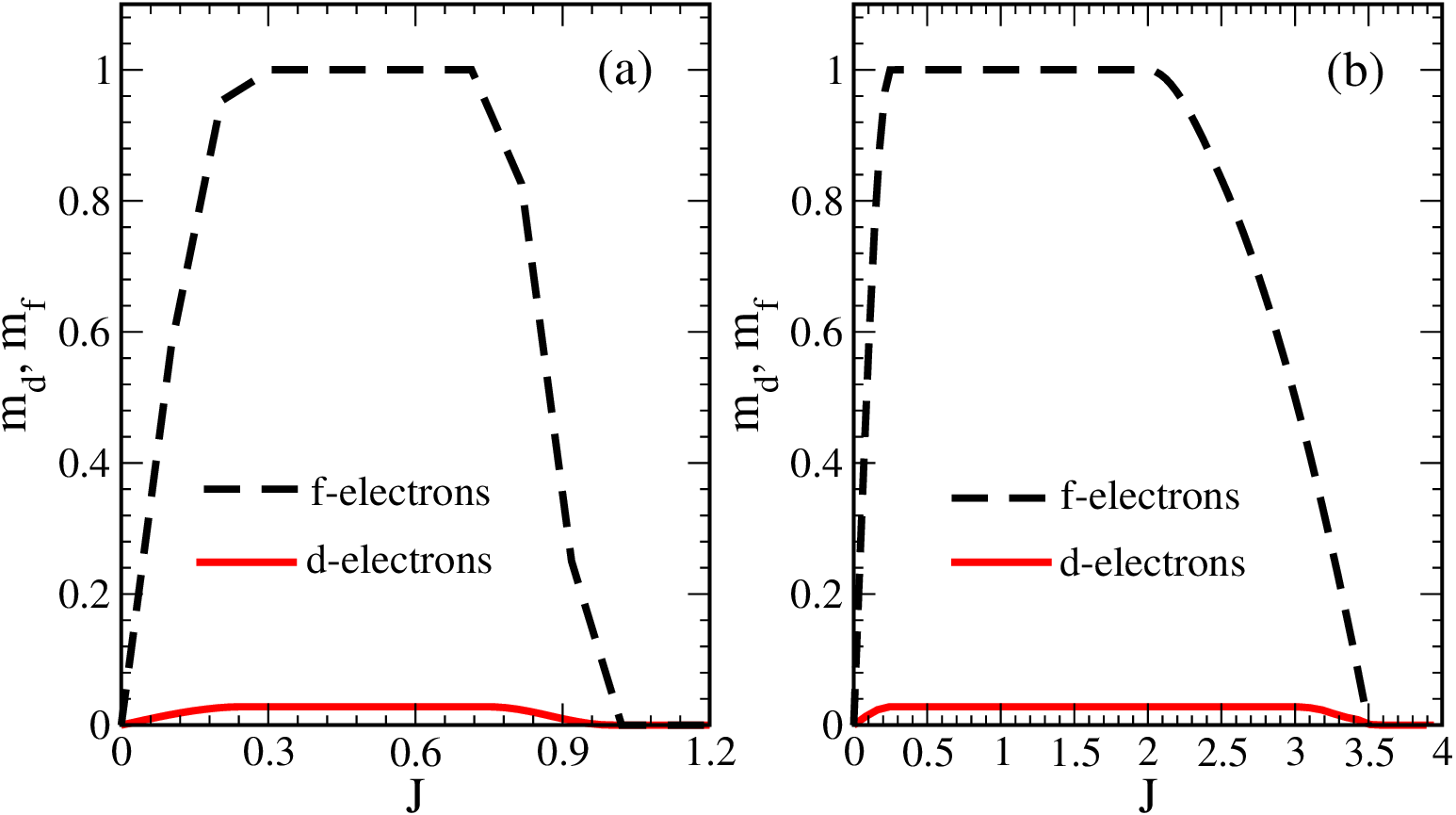}
\caption{(Color online) Variation of magnetic moment of $d$-electrons $m_{d}$ and $f$-electrons $m_{f}$ with exchange correlation $J$ for $n_{f} = \frac{1}{2}$, $n_{d} = \frac{1}{2}$ at (a) $U = 3.1$, $U_{f} = 7$  and (b) $U=5$, $U_{f} = 10$.}
\end{figure}

\begin{figure}[h]
\includegraphics[trim=0.4mm 0.4mm 0.4mm 0.1mm,clip,width=8.9cm,height=4.0cm]{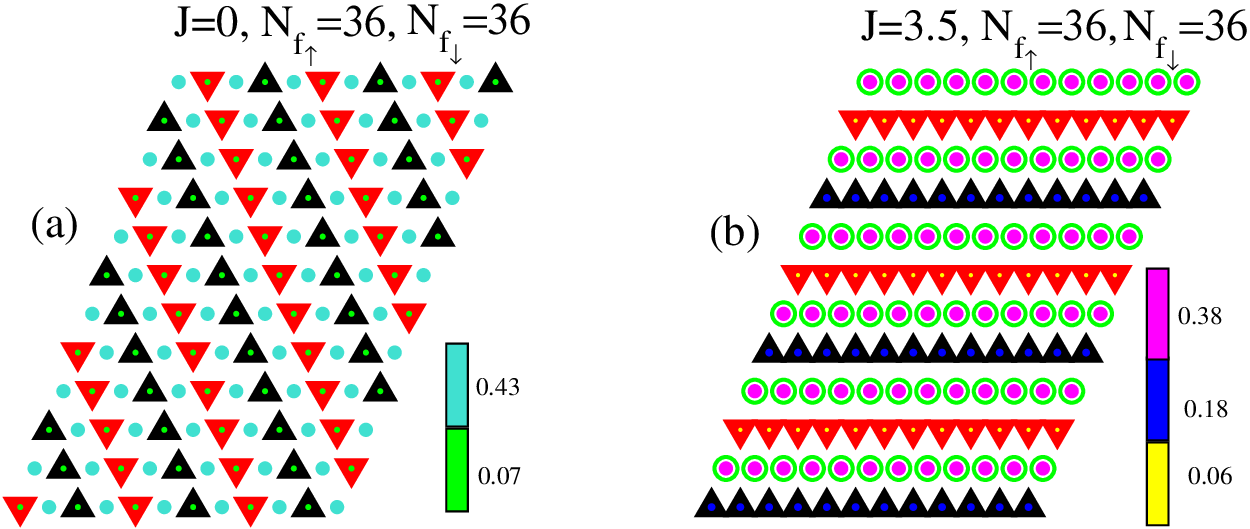}
\caption{(Color online) Up-spin $d$-electron densities are shown on each side for $J = 0$ (a) and $J = 3.5$ (b), keeping other parameters same i.e. $U = 5$, $U_{f} = 10$, $n_{f} = \frac{1}{2}$ and $n_{d} = \frac{1}{2}$. The color coding and radii of the circles indicate the $d$-electron density profile. Triangle-up and triangle-down, filled by black and red colors correspond to the sites occupied by up-spin and down-spin $f$-electrons, respectively. Open green circles correspond to the unoccupied sites.}
\end{figure}

In Fig.$3$ the ground state magnetic configurations of up-spin and down-spin $f$-electrons are shown for $U = 5$, $U_{f} = 10$ and for different $J$ values. The ground state configurations are observed to be long range ordered Neel ordered AFM arrangement of spins (Fig.3(a)) or complete FM arrangement (Fig.3(b)) or mixture of both arrangements (Fig.3(c). Here we note that $U=3.1$ is the critical value of the on-site Coulomb correlation below which we do not get finite magnetic moment. Complete AFM phases are observed below $U = 3.1$.

The variation of magnetic moment of $d$-electrons ($m_{d}$) and magnetic moment of $f$-electrons ($m_{f}$) with exchange correlation $J$ is shown in Fig.$4(a)$ and Fig.$4(b)$ for $U=3.1$ and $U_{f} = 7$ and for $U=5$ and $U_{f} = 10$ respectively. We note that in both cases the magnetic moment of $d$- and $f$-electrons increases with increasing $J$. Complete FM phase is observed at a particular value of $J$ (e.g. $J = 0.25$ for $U=5$ and $U_{f} = 10$) and it continues up to some critical value of $J$ (up to $J = 2$ for $U=5$ and $U_{f} = 10$). With further increase in $J$ mixed states are observed and finally AFM state is observed for larger values of $J$.

Figs.$5(a)$ and $5(b)$ show the density of up-spin $d$-electrons at a fixed value of $U = 5$, $U_{f} = 10$ and for $J = 0$ and $J = 3.5$ respectively. When $J = 0$ the interaction between $d$- and $f$-electrons is same irrespective of their spins so the density of $d$-electrons at sites occupied by f-electrons are same, while it is maximum at unoccupied sites. With the increase in $J$ value density of $d$-electrons at sites where $f$-electrons of same spin are present increases and at empty sites it decreases, because as $J$ increases, the interaction $(U-J)$ between $d$- and $f$-electrons of the same spins decreases.

These results can be explained as we have three kinds of competing energies in the system namely the kinetic energy of $d$-electrons and on-site interaction energies, `$U$' between $d$- and $f$-electrons of opposite spins and `$(U - J)$' between $d$- and $f$-electrons of same spins. For $J = 0$, the on-site interaction energies between $d$- and $f$-electrons of opposite spins and $d$- and $f$-electrons of same spins will be the same. Hence the ground state configuration is AFM type as possible hopping of $d$-electrons minimizes the energy of the system. It is clearly shown in the variation of $d$-electron density at each site in Fig.$5(a)$. As a result both the $m_d$ and $m_f$ are zero (as shown in Fig.$4$ and $7$). For finite but small value of $J$, the on-site interaction energy between $d$- and $f$-electrons of same spins will be smaller in comparison to the on-site interaction energy between $d$- and $f$-electrons of opposite spins. So few sites with FM arrangement of spin-up $f$-electrons will be occupied by down-spin $d$-electrons and few by up-spin $d$-electrons (as $m_d$ is very small). With this arrangement there is finite hopping possible for $d$-electrons which reduces its kinetic energy and hence total energy of the system. Therefore the ground state configuration is either mixture of AFM and FM types or completely FM type. As a result both the $m_{d}$ and $m_f$ increase with increasing $J$. Finally for large value of $J$ ($J \sim U$), the ground state is AFM in nature. We have explained that in earlier section.

\subsection{Half-filled case ($n_{f}+n_{d}=2$):}

\begin{figure}[h]
\includegraphics[trim=0.4mm 0.4mm 0.4mm 0.1mm,clip,width=8.9cm,height=4.0cm]{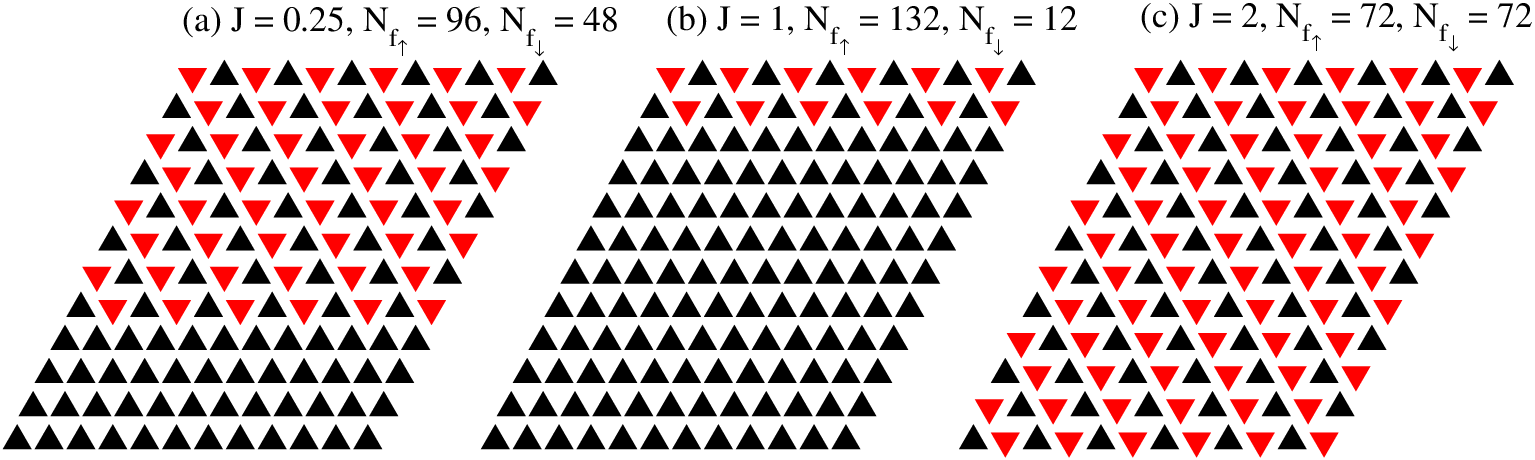}
\caption{(Color online) The ground-state magnetic configurations of $f$-electrons for $U = 5$, $U_{f} = 10$ and for various values of  $J$ with condition  $n_{f} + n_{d} = 2$. Triangle-up and triangle-down, filled by black and red colors correspond to the sites occupied by up-spin and down-spin $f$-electrons, respectively.}
\end{figure}

\begin{figure}[h]
\includegraphics[trim=0.5mm 0.5mm 0.5mm 0.1mm,clip,width=8.9cm,height=5.5cm]{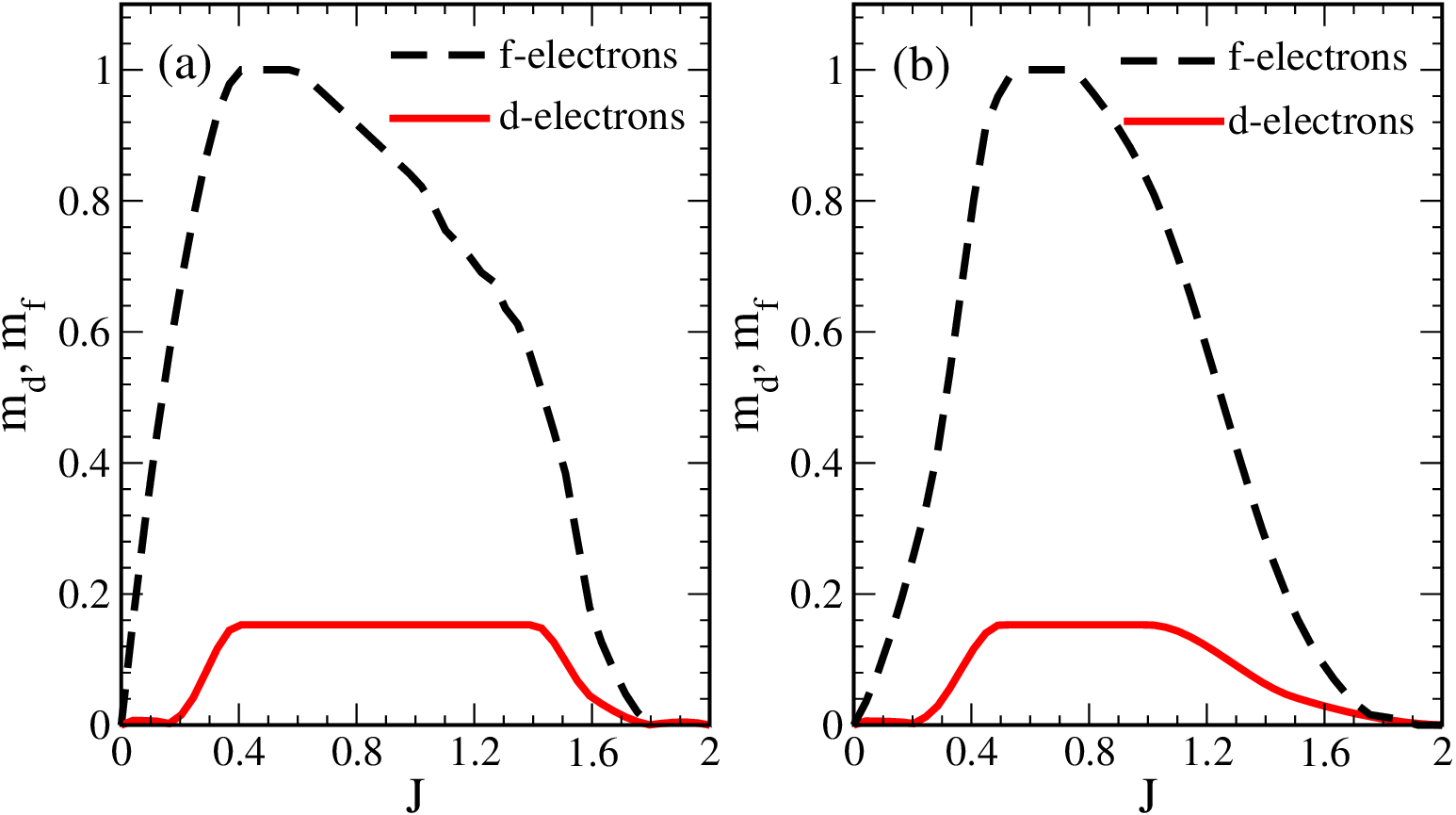}
\caption{(Color online) Variation of magnetic moment of $d$-electrons $m_{d}$ and $f$-electrons $m_{f}$ with exchange correlation $J$ for (a) $U = 2$, $U_{f}= 5$ and (b) $U = 5$, $U_{f} = 10$ and for $n_{f} = 1$ and  $n_{d} = 1$.}
\end{figure}

\begin{figure}[h]
\includegraphics[trim=0.4mm 0.4mm 0.4mm 0.1mm,clip,width=8.9cm,height=4.0cm]{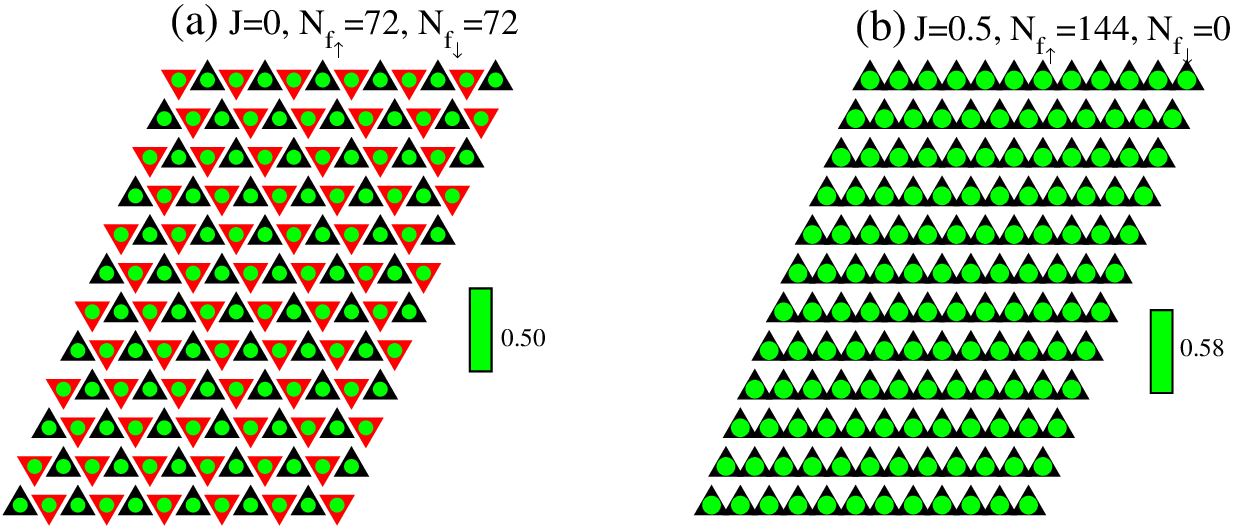}
\caption{ Up-spin $d-$electron densities are shown on each side for $J = 0$ (a) and $J = 0.5$ (b), keeping other parameters same i.e. $U = 5$, $U_{f} = 10$, $n_{f} = 1$ and $n_{d} = 1$. The color coding and radii of the circles indicate the $d$-electron density profile. Triangle-up filled by black color corresponds to the sites occupied by up-spin $f$-electrons.}
\end{figure}

Shown in Fig.$6$ are the ground state, magnetic configurations of up-spin and down-spin $f$-electrons for $U = 5$, $U_{f} = 10$ and for different $J$ values. At $J = 0$, again regular Neel ordered AFM structure is seen. With increasing $J$, mixed phase of FM and Neel type AFM is seen (Fig.6(a)). On futher increasing $J$ the phase becomes fully FM and remains the same up to a value of $J$ nearly equal to $0.75$. Then the mixed phase comes back on increasing $J$ at around $J = 1$ (Fig.6(b)) and finally the ground state becomes Neel ordered AFM again for $J > 1.80$ and continues up to $J = 5$ (Fig.6(c)). Similar FM, AFM and mixed magnetic configurations of up-spin and down-spin $f$-electrons also observed for $U = 2$, $U_{f} = 5$ and for different $J$ values (not shown here).

Corresponding variation of magnetic moment of $d$-electrons ($m_d$) and of $f$-electrons ($m_f$) with exchange correlation $J$ is shown in Fig.$7(a)$ and Fig.$7(b)$ for $U=2$ and $U_{f} = 5$ and for $U=5$ and $U_{f} = 10$ respectively. As we have already explained that the magnetic moments of $d$- and $f$-electrons increase with increasing $J$ in both cases. Increase in magnetic moment of $f$-electrons is observed when one goes from Neel type AFM to complete FM phase through the intermediate mixed phase. Similarly, $m_d$ also increases with $J$ but the increase starts after a certain value of $J$. Here we observe that $m_d$ and $m_f$ increase initially with increasing $J$ but after reaching a maximum the moments drop down at larger $J$ values and finally both $m_d$ and $m_f$ become zero. We have found that for large value of $U$, magnetization decreases sharply with $J$ in comparison to that for small values of $U$.

Figs.$8(a)$ and $8(b)$ show the density of $d$-electrons at a fixed value of $U = 5$, $U_{f} = 10$ and for  $J = 0$ and $J = 0.5$ respectively. For $J = 0$ the density of $d$-electrons at all sites is same irrespective of spins of $f$-electrons. As value of $J$ increases, density of $d$-electrons at sites where $f$-electrons of same spins are present increase as compared to at sites where $f$-electrons of opposite spins are present. This is expected because as $J$ increases, the repulsion between $d$- and $f$-electrons of the same spins decreases and hence they prefer to sit on the sites where $f$-electrons of the same spin are present.

In conclusion, the ground state magnetic properties of two dimensional spin-$1/2$ FKM on a triangular lattice for different range of values of parameters like $d$- and $f$-electron fillings, on-site Coulomb correlation $U$ and exchange correlation $J$ are studied. We have found that the magnetic moments of $d$- and $f$-electrons depend strongly on the values of $J$ and on the number of $d$-electrons $N_{d}$. We have seen for one-fourth filling that there are no magnetic moments of $d$- and $f$-electrons for $U$ less than $3.1$. At half-filling we have found that the magnetic moments of $d$- and $f$-electrons decrease sharply with $J$ at larger $U$ in comparison to smaller values of $U$. At both fillings we note that the density of $d$-electrons depends upon the value of exchange correlation $J$. Also, various charge and magnetic ordered phases of the localized $f$-electrons in the ground state have been observed at different values of $J$. The ground state configuration is observed to be long range ordered (either in some form of AFM arrangement of spins, complete FM arrangement or a mixture of both). The magnetic moments of $d$- and $f$-electrons start to increase sharply with increasing $J$ and persist up to a larger value of $J$ for one-fourth filling in comparison to one-half filling case (see Fig.4(b) and Fig.7(b)). There is rarely any calculation available for the spin-dependent FKM on a triangular lattice. Our results may motivate further studies of the magnetic properties of frustrated systems of recent interest like Cobaltates, $GdI_{2}$, $NaTiO_{2}$ and $MgV_{2}O_{4}$.

$Acknowledgments.$ SK acknowledges MHRD, India for a research fellowship. UKY acknowledges CSIR, India for a  research fellowship through SRF grant and the UGC, India for Dr. D. S. Kothari Post-doctoral Fellowship through grant No.F.4-2/2006(BSR)/13-762/2012(BSR).


\begin{thebibliography}{}
\bibitem{footnote} The large U limit of the present model would be quite interesting to look at and is clearly non trivial.
\end{thebibliography}
\end{document}